\documentclass{aa}
\usepackage{graphicx,natbib}
\usepackage{txfonts}
\begin{document}
   \title{Surface density of the young cluster \object{IC 348} in the
   \object{Perseus} molecular cloud}

   \titlerunning{\object{IC 348} surface density}

   \author{L. Cambr\'esy\inst{1} \and
	V. Petropoulou\inst{1,2} \and
	M. Kontizas\inst{2} \and
	E. Kontizas\inst{3}}

   \offprints{L Cambr\'esy\\ \email{cambresy@astro.u-strasbg.fr}} 
	   
   \institute{Observatoire Astronomique de Strasbourg, F-67000 Strasbourg,
	   France
   \and
   	Department of Astrophysics Astronomy \& Mechanics, Faculty
	of Physics, University of Athens, GR-15783 Athens, Greece
   \and
	Institute for Astronomy and Astrophysics, National Observatory
	of Athens, P.O. Box 20048, GR-118 10 Athens, Greece
   }

   \date{Received ; accepted }

   \abstract{
The \object{IC 348} young star cluster contains more than 300 confirmed members.
It is embedded in the \object{Perseus} molecular cloud, making any clustering
analysis subject to an extinction bias. In this work, we derive the extinction
map of the cloud and revisit the content of \object{IC 348} through a
statistical approach that uses the 2MASS data. Our goal was to address the
question of the completeness of \object{IC 348} and of young clusters in
general. We performed a combined analysis of the star color and density in
this region, in order to establish the surface density map of the cluster.
We reached the conclusion that \object{IC 348} has structures up to $25'$
from the cluster center, and we estimate that about 40 members brighter than
$K_s=13$~mag are still unidentified. Although we cannot use our statistical
method to identify these new members individually, the surface density map
gives a strong indication of their actual location. They are distributed in
the outer regions of the cluster, where very few dedicated observations have
been made so far, which is probably why they escaped previous identification.
In addition, we propose the existence of a new embedded cluster associated to
the infrared source \object{MSX6C G160.2784-18.4216}, about $38'$ south of
\object{IC 348}.
\keywords{Stars: pre-main sequence -- ISM: dust, extinction -- open clusters
and associations: individual: IC 348}
   }

   \maketitle

\section{Introduction}
Young stellar clusters concentrate stars in their early stages, making them
the privileged location for studying the initial conditions of star
formation from the molecular cloud fragmentation to the evolution on
the main sequence. Such a large and ambitious enterprise requires the
acquisition of multi--wavelength data from the millimeter to the
X-ray spectral range, in order to probe the different physical phenomena
that occur during star formation. The identification of cluster
members leads to constrain the cluster history, the initial mass function,
and the dynamical evolution. Although the ultimate goal is to take
a complete census of the clusters where each object is actually given a
membership status, the individual identification is not easy to obtain and
is not always necessary. A statistical identification could also lead to
information on the cluster morphology and the initial mass function.

The young cluster \object{IC 348} has been investigated regularly for
decades and is
still a very popular target. Its brightest member is the B5 star 
\object{BD +31 643}, which is part of the \object{Per OB2} association
\citep{Bla52}.
A distance of 320 pc is usually adopted for the cluster although it is
still controversial \citep{Cer93,Her98,dHd+99}. \cite{Her54} identified 16
members characterized by their $H\alpha$ emission, and the current number 
is about 300, with a mass and age distribution peak at 0.2 M$_\odot$
and 2 Myr, respectively \citep{Luh99,Car02,LSM+03,MLL+03,PSZ03,PZ04,LMG05}.
\citet{MLL+03} showed that the extinction reaches values as high as 20 mag
toward
the cluster line of sight, the cluster itself being only partially embedded
in the \object{Perseus} molecular cloud with an average extinction of 4.5 mag
\citep{LL95}. The extinction will introduce a bias in the apparent member
distribution, as long as the cluster census is incomplete.
It is a general characteristic of young clusters to
still be associated with their parent molecular cloud. The apparent stellar
density, which makes a cluster detectable, results from both the true cluster
density and the extinction variation toward its direction. Although a classical
clustering analysis is biased in this configuration, it is possible to
extract relevant information on clusters with a statistical analysis which
takes the extinction into account \citep{Car00,CBJC02}.


\section{Cluster mapping}\label{method}
\begin{figure}
  \centering
  \includegraphics[width=8.8cm]{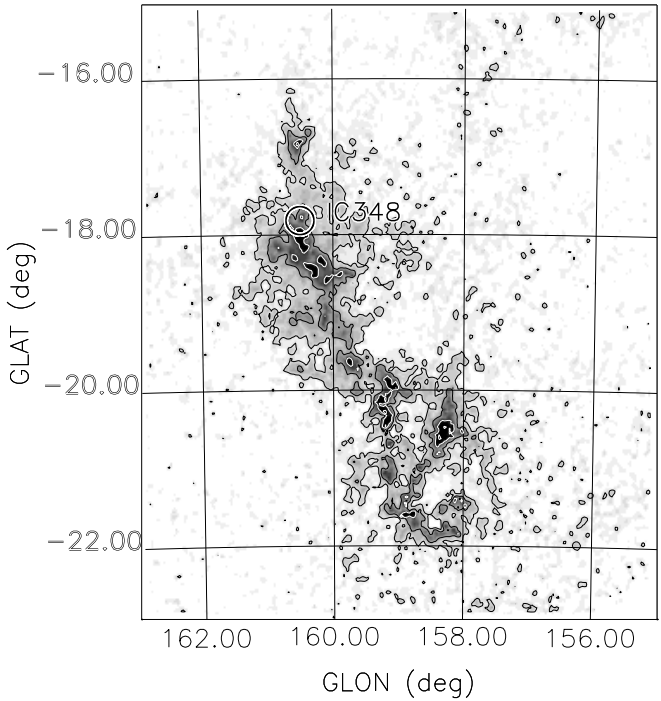}
  \caption{\object{Perseus} extinction map from 2MASS $H-K_s$ color excess.
  The spatial resolution is $4'$ and isocontours are for $A_V=$2, 4, 8, 16
  mag. The circle position and radius are set to contain all the
  confirmed members of \object{IC 348}.}
  \label{avmap}
\end{figure}
The \object{Perseus} molecular cloud has an elongated and filamentary
structure, as shown in the extinction map presented in Fig. \ref{avmap}. This
map is obtained by measuring the 2MASS $H-K_s$ color excess in cells with
adaptive size. The cell size is chosen to contain only 3 stars, and the full
map is then convolved by an adaptive kernel to obtain a final uniform spatial
resolution of $4'$. This technique reduces the non-linear error because of the
extinction variations at a smaller scale than the map resolution \citep{CJB05}.
The known members of \object{IC 348} were removed before the extinction map
was built.

The final map is very similar to the $^{13}$CO map \citep{PBB+99} and to the
extinction map presented on the COMPLETE 
webpage\footnote{http://www.cfa.harvard.edu/COMPLETE/} with the NICER method
\citep{LA01}.
An extinction estimate from reddening or from star counts relies on the
assumption that all stars are behind the obscuring cloud and that the
stellar population is homogeneous over the whole field.
When a young stellar cluster is embedded in a molecular cloud, none of
the conditions is fulfilled. The color and the star count maps react
differently to the cluster contamination. \citet{CBJC02} proposed using
this property to identify clusters by comparing both maps. The next 3
sections describe the technique to build the color map and the count map
and to obtain the cluster morphology.

In the following, we focus on a $2\degr \times 2\degr$ field centered on IC
348 at glon$=160.38\degr$, glat$=-17.78\degr$.

\subsection{Reddening}
The technique is based on the pioneering work by \citet{LLCB94}, who propose a
statistical method to build extinction maps from near--infrared color
excesses. The specificity of our method is to use adaptive cells that
contain a fixed number of stars, in order to increase the spatial resolution
and to limit the bias due to the inhomogeneous distribution of stars that
tend to concentrate in the less obscured part of a cell. A detailed
description of this bias can be found in \citet{CJB05}. The visual
extinction was obtained through the 2MASS $H-K_s$ color excess as follows:
\begin{eqnarray}
\nonumber
E_{H-K_s} &=& (H-K_s)_{\rm obs}-(H-K_s)_{\rm int}\\
A_V^{\rm col} &=& \left(\frac{A_H}{A_V} - \frac{A_{K_s}}{A_V}\right)^{-1}
		E_{H-K_s}\label{eq_col}
\end{eqnarray}
where the indices $obs$ and $int$ stand for $observed$ and $intrinsic$,
respectively. Here, $(H-K_s)_{\rm int} = 0.12 \pm 0.1$ mag from the region
outside the \object{Perseus} cloud.
We used the extinction law from \citet{RL85}, for which $A_H/A_V=0.175$ and
$A_{K_s}/A_V=0.112$, yielding the color excess to visual extinction conversion
formula $A_V=15.87 \times E_{H-K_s}$.
The presence of a young embedded cluster will affect the extinction, because
1) YSOs are embedded in the cloud and suffer only a partial obscuration, and 
2) YSOs may have different colors than field stars, with an excess at $K_s$ due
to an accretion disk for those objects that are younger than about 2 Myr. In
the former case, the extinction will be underestimated, in the latter it
will be overestimated. 
The effect does not usually stand out when examining the reddening map
because no anomaly (e.g. strong gradient or discontinuity) appears there.
Another independent extinction estimation is required to detect the cluster
contamination.

\subsection{Star counts}
In the star count method, the decrease in the apparent stellar density is
interpreted as a foreground obscuration.
It is the historical method for mapping the interstellar extinction
\citep{Wol23,Bok56}. As mentioned above, we used an adaptive technique with
the exact same cells as for the color excess mapping. The visual extinction is
obtained by assuming the apparent variations of the stellar density are due
to foreground obscuration:
\begin{eqnarray}
A_{K_s}^{\rm cnt} &=& \frac{1}{a} \log \frac{D_{\rm ref}}{D}\label{eq_cnt}\\
\nonumber
A_V^{\rm cnt}     &=& \frac{A_V}{A_{K_s}} \times A_{K_s}^{\rm cnt}
\end{eqnarray}
where $a$ is the slope of the $K_{s}$ luminosity function (KLF), $D$ the
star density in each cell, and $D_{\rm ref}$ the density in a reference
field. Although the reference star density formally depends on the position
(mainly on the galactic latitude), a constant value is correct for small
areas of about 1 deg$^2$. When a cluster contaminates the field, the stellar
density is locally enhanced, and the extinction is biased toward low values.
If the local density is higher than the average density over the field
($D_{\rm ref}$), which is true if the extinction is small, it gives a
negative number for the extinction. Obviously, Eq. \ref{eq_cnt} is no longer
correct, because the measured density actually includes the background stars
and the cluster stars : $D = D_{\rm b}+D_{\rm cl}$

\subsection{Comparison of the two maps}
Comparison of the maps requires the two calibrations to be consistent.
The spatial resolution is identical in both maps by construction.
The relative calibration of the two maps is obtained using an arbitrary
value for $D_{\rm ref}$ in Eq. \ref{eq_cnt} and by forcing
the median of $\Delta A_V=A_V^{\rm col} - A_V^{\rm cnt}$ to zero.
The absolute calibration of the extinction maps is not relevant for
the cluster study.

The difference $\Delta A_V$ should be a structureless map, if the field is
free of any cluster contamination.
Otherwise, the cluster morphology will be revealed by the structures
contained in the $\Delta A_V$ map because $D_{\rm cl} > 0$, meaning
$A_V^{\rm col} > A_V^{\rm cnt}$.


\section{Application to the \object{IC 348} young cluster}\label{ic348}
\subsection{Data}
\begin{figure}
  \centering
  \includegraphics[width=8.8cm]{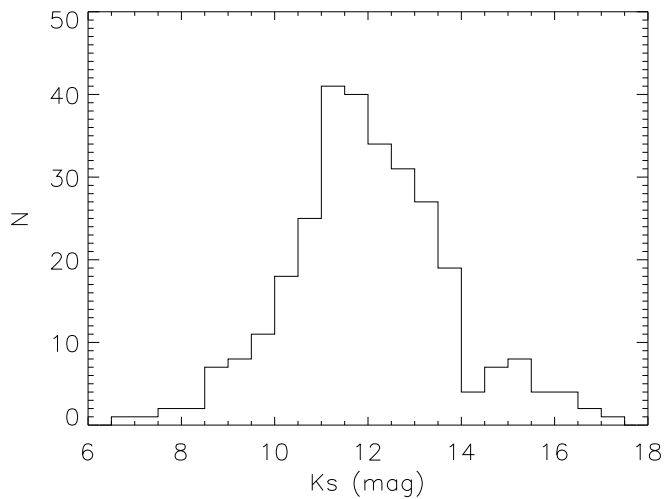}
  \caption{$K_s$ luminosity function of the 297 stars of the cluster with
  available $K_s$ magnitude \citep{LSM+03,LLME05}.} 
  \label{klf}
\end{figure}
The data were extracted from the 2MASS All-Sky Point Source Catalog
\citep{CSv+03} for the region defined by $158.6\degr<{\rm glon}<162.2\degr$
and $-19.2\degr <{\rm glat}<-16.3\degr$.
The list of the 304 confirmed members of \object{IC 348} was provided by
\citet{LSM+03} and \citet{LLME05}, 278 of which are present in 2MASS.
Figure \ref{klf} shows the $K_s$ luminosity function of the full member list.
The distribution peaks at $K_s\approx 11.5$~mag with more than 75\% of the
sources brighter than $K_s=13$~mag. On the other hand, faint stars dominate
the background star population with a number of sources proportional to 
$10^{a\times K_s}$, with $a=0.31$. While a magnitude cut at $K_s<13$~mag keeps
75\% of the cluster members, it removes 75\% of the background stars.
In order to increase the signal-to-noise ratio in the $\Delta A_V$ map,
we applied this cut before building the reddening and star count maps.

\subsection{Results}
\begin{figure}
  \centering
  \includegraphics[width=8.8cm]{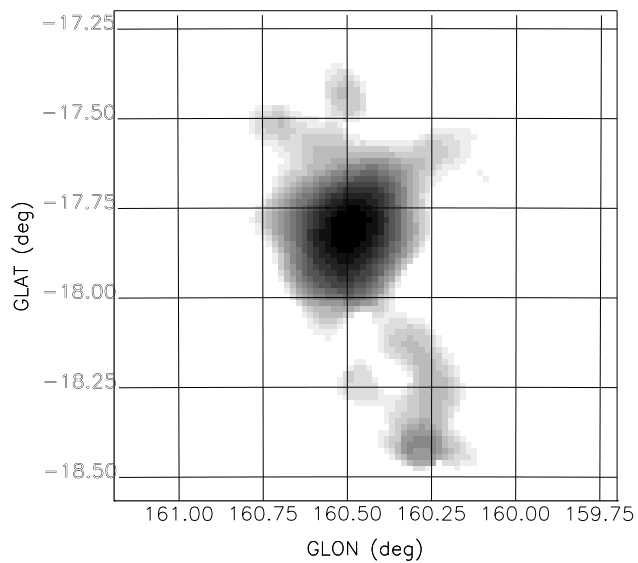}
  \caption{Surface density map of \object{IC 348} obtained from stars brighter
  than $K_s=13$~mag. The map shows the region $2 \sigma = 7.7$~mag
  above the noise.}
  \label{cluster}
\end{figure}

Following the method of Sect. \ref{method}, the extinction was from
adaptive cells that contain 3 stars. The reddening and the count maps
were then smoothed with a Gaussian kernel to reach the final uniform resolution
of $8'$. We stress that it requires a convolution by an adaptive
kernel with ${\rm FWHM}=({8^2 - {\rm FWHM}^2_{\rm orig}})^{1/2}$, where
${\rm FWHM}_{\rm orig}$ is the resolution in arcmin of the original maps.
The difference between the maps from Eqs. \ref{eq_col} and \ref{eq_cnt}
is presented in Fig. \ref{cluster}. As expected a peak appears at the
\object{IC 348} position.

\begin{figure}
  \centering
  \includegraphics[width=8.8cm]{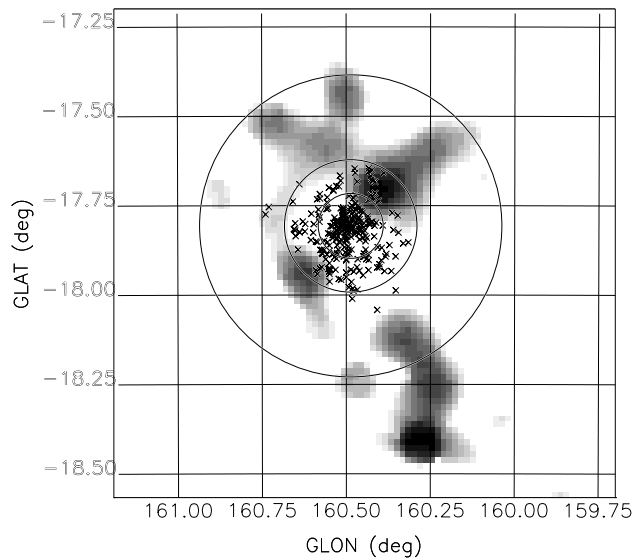}
  \caption{Same as Fig. \ref{cluster} but derived from a catalog filtered from
  all known \object{IC 348} members. These members are overlaid on the map as
  crosses. Circles at $5'$, $10'$, and $25'$ radius define the {\em core}
  \citep{LL95}, the {\em halo} \citep{MLL+03}, and the cluster boundary
  suggested from our study.}
  \label{sans_cluster}
\end{figure}

The very same process was repeated on the 2MASS catalog filtered from all
known \object{IC 348} members. The resulting $\Delta A_V$ map is presented
in Fig. \ref{sans_cluster} with the cluster members overlaid as crosses.
Significant structures remain on the map, which strongly supports the idea
that the cluster extends beyond the {\em halo} defined by \citet{MLL+03} at
$10.33'$ and beyond the ROSAT X-ray detection boundary at $15'$ \citep{PZH96}.
The map exhibits a faint ring around the central part of the cluster and
several substructures. The farthest clump, located at $l=160.26$ $b=-18.42$,
is $38'$ south of the \object{IC 348} center and seems linked
to the main cluster by a ridge. It contains an infrared source reported
by IRAS and MSX, \object{MSX6C G160.2784-18.4216} \citep{KSP+03}. It is,
however, likely that this structure corresponds to another star-formation
site in \object{Perseus}. Further analysis needs to be carried out to conclude
whether it is related to IC 348. Another clump located to the north-west of
\object{IC 348} has the particularity of being within the {\em halo} radius 
\citep{MLL+03} of \object{IC 348}, which means this region has already
been studied, although several bright members were not identified.  

Figures \ref{cluster} and \ref{sans_cluster} were obtained by subtracting two
extinction maps, making the pixel unit to be {\em magnitude}. It has no
physical meaning in this context, except that it goes with the star density.
A straightforward way to convert these magnitudes into star density would
require the real extinction to be known, but both the reddening and the
count maps are contaminated by the YSOs.
Fortunately, the outer region of \object{IC 348} has a low star density,
since it is the first time the excess has been detected. It means the sample
must be dominated by background stars, which makes the $H-K_s$ color excess
an acceptable estimation of the extinction. We stress that the median color
is used within cells rather than the mean, which ensures a low sensitivity
to the YSOs contamination as long as it is less than 50\%. Assuming the
color-based extinction map is correct, Fig. \ref{sans_cluster} was converted
in surface density as follows:
\begin{equation}
D_{\rm cl} = c \times \left(10^{a\; \Delta A_{K_s}} -1 \right)\label{eq.nbyso}
\end{equation}
where $c$ is obtained using the knowledge that the difference between
$N_{\rm cl}$ in Figs. \ref{cluster} and \ref{sans_cluster} must be exactly
the number of confirmed members that were removed from the original 2MASS
catalog to make the second figure (i.e. 212 stars for $K_s<13$~mag). We
finally conclude that $40 \pm 6$ stars with $K_s<13$~mag are responsible for
the excess seen in Fig. \ref{sans_cluster}. The quoted uncertainty is the
statistical uncertainty associated with the Poisson law followed by the star
distribution. A total number of 350 stars brighter than 13~mag at $K_s$ was
found in the area, meaning an 11\% level contamination by the cluster, which
justifies our choice of relying on the color median as an extinction estimator.
Following Eq. \ref{eq.nbyso}, the southern clump centered on
\object{MSX6C G160.2784-18.4216} would contain about 10 YSOs
brighter than $K_s<13$~mag.

\begin{figure}
  \centering
  \includegraphics[width=8.8cm]{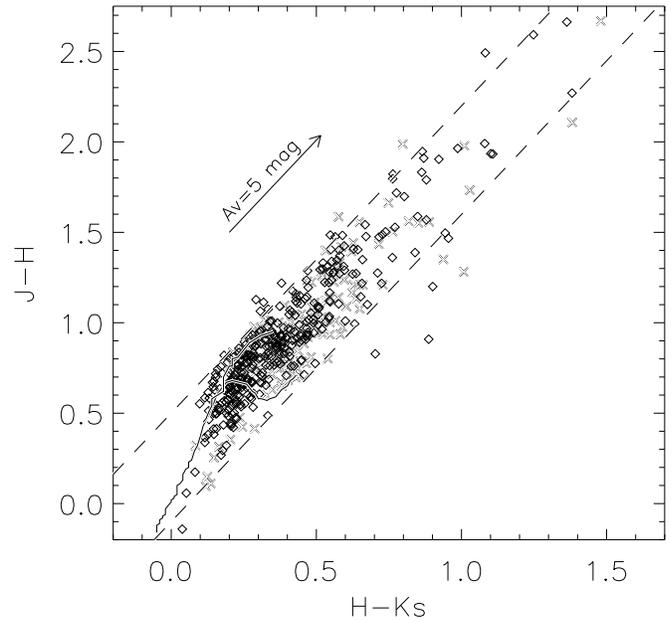}
  \caption{Color-color diagram for stars with $K_s<13$~mag in the area
           defined by Fig. \ref{sans_cluster} (diamonds) and for confirmed
	   \object{IC 348} members (crosses). The dashed lines are parallel
	   to the reddening vector.}
  \label{colcol}
\end{figure}

The color-color diagram (Fig. \ref{colcol}) for the stars located in the
region defined in Fig. \ref{sans_cluster} shows almost no star
with a $H-K_s$ excess. This is also true for the confirmed \object{IC 348}
members, which indicates that most YSOs have already lost their inner
accretion disk that is responsible for the infrared excess at 2 $\mu$m.
Individual identification is therefore not possible with the 2MASS photometry.


\section{Conclusions}
The statistical comparison of the stellar color and density permit building
the surface density map of embedded clusters. We applied our technique to
\object{IC 348}, using two different sample catalogs, one with all 2MASS
sources and the other one filtered from all known associated YSOs. As
expected, the first map shows a peak at the cluster location, but it is
more surprising to find evidence of structures in the filtered map, too. 
These structures are distributed around \object{IC 348}, suggesting they are
actually an extension of about $40$ new members for the well-known cluster.
This result encourages more detailed study of the structure of \object{IC 348}
and, in particular, of its dynamical evolution, as it created these
clumps in the cluster. As no infrared excess was found in 2MASS, we cannot 
identify the potential new members individually. Longer wavelength follow-up
with Spitzer may exhibit the presence of a residual disk, but X-ray
observations with XMM-Newton are probably more appropriate to finding the older
pre-main-sequence stars in the outer region of \object{IC 348}. Spectroscopic
follow-up remains the most suitable way to obtain the final diagnostic for
membership. Proper motion measurements could be an indicator of the
membership and would be of great interest for the kinematic study
of the cluster.

\acknowledgement
This publication makes use of data products from the Two Micron All Sky Survey, which is a joint project of the University of Massachusetts and the Infrared Processing and Analysis Center/California Institute of Technology, funded by the National Aeronautics and Space Administration and the National Science Foundation.


\end{document}